\documentstyle[12pt,preprint]{aastex}
\begin{document}

\title{Early Results from the Galactic O-Star Spectroscopic Survey: 
C~III Emission Lines in Of Spectra}

\author{Nolan R.\ Walborn,\altaffilmark{1} Alfredo Sota,\altaffilmark{2} 
Jes\'us Ma\'{\i}z Apell\'aniz,$^{2,3}$ Emilio J. Alfaro,\altaffilmark{2}\\ 
Nidia I.\ Morrell,\altaffilmark{4} Rodolfo H.\ Barb\'a,$^{5,6}$ Julia I.\ Arias,\altaffilmark{5} \& Roberto C.\ Gamen\,\altaffilmark{7}}

\altaffiltext{1}{Space Telescope Science Institute,$^{*}$ 3700 San Martin Drive, Baltimore, MD 21218; walborn@stsci.edu}

\altaffiltext{2}{Instituto de Astrof\'{\i}sica de Andaluc\'{\i}a--CSIC, Glorieta de la Astronom\'{\i}a s/n, 18008 Granada, Spain; sota@iaa.es, jmaiz@iaa.es, emilio@iaa.es}

\altaffiltext{3}{Ram\'on y Cajal Fellow}

\altaffiltext{4}{Las Campanas Observatory, Observatories of the Carnegie Institution of Washington, Casilla 601, La Serena, Chile; nmorrell@lco.cl}

\altaffiltext{5}{Departamento de F\'{\i}sica, Universidad de La Serena, Cisternas 1200 Norte, La Serena, Chile; rbarba@ dfuls.cl, julia@dfuls.cl}

\altaffiltext{6}{Also Instituto de Ciencias Astron\'omicas de la Tierra y del Espacio (ICATE--CONICET), Avenida Espa\~na 1512 Sur, J5402DSP, San Juan, Argentina}

\altaffiltext{7}{Instituto de Astrof\'{\i}sica de La Plata--CONICET and Facultad de Ciencias Astron\'omicas y Geof\'{\i}sicas, Universidad Nacional de La Plata, Paseo del Bosque s/n, 1900 La Plata, Argentina; gamen@fcaglp.unlp.edu.ar}

\altaffiltext{*}{Operated by the Association of Universities for Research in Astronomy, Inc., under NASA contract NAS5-26555} 

\begin{abstract}
On the basis of an extensive new spectroscopic survey of Galactic O~stars, 
we introduce the Ofc category, which consists of normal spectra with C~III 
$\lambda\lambda$4647--4650--4652 emission lines of comparable intensity to 
those of the Of defining lines N~III $\lambda\lambda$4634--4640--4642.  The former feature is strongly peaked to spectral type O5, at all luminosity 
classes, but preferentially in some associations or clusters and not others.  The relationships of this phenomenon to the selective C~III $\lambda$5696 emission throughout the normal Of domain, and to the peculiar, variable Of?p category, for which strong C~III $\lambda\lambda$4647--4650--4652 
emission is a defining characteristic, are discussed.  Magnetic fields have 
recently been detected on two members of the latter category.  We also present two new extreme Of?p stars, NGC~1624--2 and CPD~$-28^{\circ}$~2561, bringing the number known in the Galaxy to five.  Modeling of the behavior of these spectral features can be expected to better define the physical parameters of both normal and peculiar objects, as well as the atomic physics involved. 
\end{abstract}


\section{Introduction}

Comprising the most massive and energetic stars in the Galaxy, the Population~I O~spectral class is of considerable interest and has been the subject of many investigations.  The optical spectra of over 1000  representatives have been observed.  Their detailed spectral classification was begun by Plaskett \& Pearce (1931), who defined subtypes O5--O9 on the basis of helium-line ionization ratios, and denoted as Of the subset with selective emission lines of He~II $\lambda$4686 and N~III $\lambda\lambda$4634--4640--4642.  This work was incorporated into the MK System (Johnson \& Morgan 1953; Abt, Meinel, Morgan, \& Tapscott 1968; Morgan \& Keenan 1973), with the addition of luminosity classes V--I at subtypes O8--O9, and eventually of the earlier subtype O4.  Further developments with higher photographic resolutions were introduced by Walborn (1971a, 1973a), Conti \& Alschuler (1971), and Conti \& Leep (1974), including luminosity classes at all subtypes, with the Of stars interpreted as the normal supergiants.  The still earlier subtypes O3 and O2 were added by Walborn (1971b) and Walborn et~al.\ (2002), respectively, and a digital atlas was provided by Walborn \& Fitzpatrick (1990).  The complete historical and technical details of OB spectral classification have recently been reviewed by Walborn (2009).

Why then should further spectral classification work on the known Galactic O~stars be undertaken?  A major reason is the substantially superior quality and speed of modern CCD detectors, which render a complete, homogeneous, and higher S/N survey feasible.  Such a study may be expected to improve the systematic and random accuracy of the classifications, while revealing new categories and peculiar objects within the sample.  Moreover, further information about the endemic multiplicity of the O~stars is essential for correct inference of their physical properties and evolution.  Accordingly, we are conducting a massive digital survey of the optical spectra of all Galactic O~stars in both hemispheres accessible to our available equipment as detailed below (Galactic O-Star Spectral Survey, GOSSS; J.M.A., overall/northern PI; R.H.B.\ and N.I.M., southern Co-PIs), i.e., to about 13th magnitude and 1300 stars.  The first installment of the survey, including a new spectral classification atlas, will be presented by Sota et~al.\ (2010).  Eventually, the data for all the stars will be linked to the public online Galactic O-Star Catalogue (GOSC; Ma\'{\i}z Apell\'aniz et~al.\ 2004; http://dae45.iaa.csic.es:8080/$\sim$jmaiz/research/GOS/GOSmain.html).
Here we present some initial examples of the unexpected developments that 
such a survey can produce.

\section{Observations}

The data have been acquired at three observatories and have similar albeit not identical characteristics.  At the Observatorio de Sierra Nevada (OSN, Spain), the 1.5~m reflector and Albireo spectrograph with an 1800~l~mm$^{-1}$ grating provide a dispersion of 0.66~\AA~pixel$^{-1}$. At the Centro Astron\'omico Hispano Alem\'an (CAHA, also Calar Alto Observatory, Spain), the 3.5~m reflector and TWIN spectrograph with a 1200~l~mm$^{-1}$ grating provide a dispersion of 0.54~\AA~pixel$^{-1}$. At the Las Campanas Observatory (LCO, Chile), the 2.5~m Du Pont telescope and Boller \& Chivens spectrograph with a 1200~l~mm$^{-1}$ grating provide a dispersion of 0.80~\AA~pixel$^{-1}$.  The spectral resolving power is 
$\sim$3000 in all three cases, and the typical S/N is 200--300.  Further details about the data and reductions will be presented by Sota et~al.\  (2010).  The stars and observations to be discussed here are listed in Table~1, while segments of the spectrograms are displayed in Figures~1 and 2 in the same order.

\section{Results}

\subsection{The Ofc Stars}

A luminosity sequence of the new Ofc spectral category is shown in Figure~1.  Although there were previous indications in some spectra (Walborn 1973b, 
Herrero et~al.\ 1999, Walborn \& Howarth 2000), we were initially surprised 
to find strong C~III $\lambda\lambda$4647--4650--4652 emission lines in our
data for four high-luminosity early-O stars belonging to the Cygnus~OB2 
association.  However, as study of our large sample of high-S/N data 
proceeded, further examples of this phenomenon were encountered in other 
regions and at all luminosity classes, but always at or near spectral type O5, 
as seen in Figure~1.  As shown by Walborn \& Fitzpatrick (1990; their Figs.~4 
\& 5) and Sota et~al.\ (2010), although often present, the C~III is much weaker than the N~III at earlier and later spectral types.  There are also some O4--O5 spectra in which, although still strong, the C~III is weaker than 
the adjacent N~III lines, e.g., in HD 15570, O4~If and HD~14947, O5~If. In some O4 spectra, the C~III emission lines are joined by C~IV $\lambda$4658 
emission of comparable intensity, e.g., HD~93250, O4~III(fc) and HD~15558, 
O4.5~III(fc); cf. Cyg~OB2--9 in Fig.~1 and see Walborn et~al.\ 2002 for
identification of the C~IV line in O2--O3 spectra.  

From investigation of the first $\sim$300 spectra in our survey, 18 of types 
O3.5-O5.5 have been classified as Ofc so far.  Of these, four are in Cygnus~OB2, eight in the Carina Nebula complex, and two in IC~1805; while one each are the only (optically bright) early-O stars in NGC~1893, NGC~6334, and NGC~6530.  This subsample of our survey also contains 21 O4--O5 spectra not classified as Ofc; however, five of these are marginal cases with strong C~III emission somewhat weaker than the N~III, including two in the Carina Nebula and one in IC~1805.  Others of these non-Ofc types are peculiar objects (five Onfp, two O~Iafpe).  But two of them are the primary classification standards HD~46223, O4~V((f)) and HD~46150, O5~V((f)) in NGC~2244, and another is the brightest early-O star in NGC~6611.  These demographics suggest that a spectral type near O5 is a necessary but not sufficient condition for the Ofc phenomenon, which occurs primarily in certain associations and young clusters, but not others.  Thus, a cluster parameter in addition to the stellar spectral type plays a role, such as initial rotational velocities (Maeder \& Meynet 2000), age, and/or chemical composition. These full samples and future ones will be presented and further
discussed by Sota et~al.\ (2010) and in subsequent papers.   

Some revisions and refinements of spectral categories and individual types will be noted in this discussion.  Along with the recognition of the Ofc category, these are developments resulting from the quality and extent of our new survey data; they will be fully described and a comprehensive new spectral atlas will be presented in Sota et~al.\ (2010). For instance, the ``+'' parameter previously used to denote Si~IV $\lambda\lambda$4089--4116 emission in O-type spectra has been dropped, in view of the expanded array of selective emission lines that are now being recognized (Walborn 2001; Werner \& Rauch 2001; Walborn et~al.\ 2004a; Corti, Walborn, \& Evans 2009).  However, it has been deemed useful to add the ``c'' to emphasize C~III emission in the new category discussed here, because of the prior definition of the Of phenomenon in terms of the adjacent N~III and He~II $\lambda$4686 emission lines only.  The double and single parentheses introduced by Walborn (1971a) to describe weak N~III emission in combination with strong He~II absorption, and stronger N~III with the He~II weakened or neutralized, respectively, are retained in the Ofc types; the unadorned ``f'' or ``fc'' is reserved for all these features strongly in emission, and this progression is now recognized as a luminosity effect in normal spectra, with the last case corresponding to the supergiants. 

\subsection{The Of?p Stars}

The Of?p notation was introduced by Walborn (1972) to describe two peculiar spectra (HD~108 and HD~148937) and emphasize doubt that they
are normal supergiants, as the normal Of spectra had just been interpreted.  A third example of the category (HD~191612) was discovered by Walborn (1973a).  This distinction has been amply vindicated by subsequent developments.  First, extreme, recurrent spectral variations were found in HD~108 (Naz\'e, Vreux, \& Rauw 2001) and HD~191612 (Walborn
et~al.\ 2003); also, as shown in the latter paper, these stars do not have
supergiant wind profiles in the ultraviolet.  A period in excess of 50~yr is 
indicated for HD~108 (Naz\'e et~al.\ 2001), whereas a strict period of 538~d 
was identified in HD~191612 (Walborn et~al.\ 2004b; Howarth et~al.\ 2007); these are now believed to be rotational periods.  Then, HD~191612 (Donati et~al.\ 2006) and HD~108 (Martins et~al.\ 2010) have become two of the first few O~stars with detected magnetic fields, which most likely reveal their basic nature of braked, oblique rotators. 

The most prominent defining characteristic of the Of?p category was the
presence of C~III $\lambda\lambda$4647--4650--4652 emission lines of
comparable intensity to the N~III $\lambda\lambda$4634--4640--4642.
However, it is important to note that there were others, including sharp
absorption, emission, and P~Cygni features at H and He lines indicative of 
circumstellar structures, and a general lack of other Of supergiant properties such as prominent Si~IV absorption lines and broad emission pedestals beneath the narrow emission lines.  Now, as shown in the references cited and Figure~2, it has been found that the C~III emission even disappears entirely at the late-type or ``minimum'' phases of HD~108 and HD~191612! It was fortuitous that both of these stars were near their maxima during the observations by Walborn (1973a).

Figure~2 shows a key segment in the spectra of the three previously known
Galactic Of?p stars (with HD~108 appearing in its current minimum state and 
HD~191612 near both extremes of its variability range), together with two 
newly recognized similar objects that extend the phenomenology of the category.

The star NGC~1624--2 is the brightest in the ionizing cluster of the small, 
distant anticenter H~II region Sh2--212 (Sharpless 1959; Moffat, FitzGerald, 
\& Jackson 1979).\footnote{Currently this object can be identified in SIMBAD 
either as MFJ~Sh~2--212~2 or as GSC~03350--00255.  Although SIMBAD reports them as different objects separated by 1\farcs4, the two identifiers refer to the same star; SIMBAD further confuses the issue by calling the latter 
NGC~1624--1, as do Chini \& Wink (1984).  See Deharveng et~al.\ (2008) for 
a definitive identification chart, as well as an interesting study of the region.  NGC~1624--2 is also 2MASS~04403728+5027410.} We have found that it has an extreme, so far unique Of?p spectrum, with the C~III emission blend substantially {\it stronger\/} than the N~III features.  The two-dimensional spectral images show that the strong Balmer emission is definitely circumstellar and not from the H~II region; a few repeated observations indicate that it moves with respect to the underlying absorption 
lines (or vice versa), with a small amplitude and a timescale of days, but 
further observations are ongoing to better characterize this behavior.  

CPD~$-28^{\circ}$~2561 was already classified as a peculiar Of object of
undetermined nature by Walborn (1973a) and Garrison, Hiltner, \& Schild
(1977).  Repeated high-resolution observations, associated with the GOSSS
for the investigation of spectroscopic binaries (in the south, the Galactic O and WN [OWN] Survey, R.C.G.\ and R.H.B., Co-PIs), now show that this star undergoes extreme spectral transformations very similar to those of HD~191612, on a timescale of weeks (Barb\'a et~al., in preparation). The C~III $\lambda\lambda$4647--4650--4652 emission intensity has not yet been seen to rival that of N~III $\lambda\lambda$4634--4640--4642, but it is definitely variable and the maximum may not yet have been covered.

\section{Discussion}

Selective emission lines in O-type spectra, i.e., lines that come into emission while others from the same ions remain in absorption (Walborn 2001) are potentially a very powerful but as yet poorly developed diagnostic of physical conditions in the atmospheres of these stars, as well as of anomalous 
level populations in those ions.  It is well established that these emission lines display smooth correlations with the stellar spectral types and luminosity classes, and that they have the same profiles as the stellar absorption lines in normal spectra, i.e., they arise in the stellar photospheres.  However, the only one of the growing array of these features 
that has been subjected to detailed theoretical investigation in the literature 
is the N~III $\lambda\lambda$4634--4640--4642 triplet, long ago by Mihalas, 
Hummer, \& Conti (1972) and Mihalas \& Hummer (1973).  Corti et~al.\ (2009) provided preliminary evidence that current state-of-the-art atmospheric models can reproduce these lines, or not, depending upon quite small changes in the basic parameters, but an effort to do so systematically and understand the atomic physics involved is as yet lacking.

The behavior of the C~III $\lambda\lambda$4647--4650--4652 triplet in the 
Ofc spectra reported here constitutes a new selective emission effect, which 
contrasts sharply with the well-known C~III $\lambda$5696 singlet emission.  
While the former is very sharply peaked at spectral type O5, the latter pervades nearly the entire O-type domain (Conti 1974; Walborn 1980).  The
triplet absorption vs.\ singlet emission dichotomy in these C~III lines was 
already described by Swings \& Struve (1942) in the spectra of massive 
O~stars, while Smith \& Aller (1969) and Heap (1977) discussed analogous 
effects in Of central stars of planetary nebulae.  Physical explanations of these distinct behaviors will be of considerable interest.  They are unlikely to be solely chemical abundance effects, because the spectral types are based on He ionization ratios, and the luminosity classes on the absolute strengths of the He~II, N~III, and as of now in some spectra near O5, C~III emission lines.  Nevertheless, there may be a secondary role of abundances, whether initial or from mixing of CNO-cycled material in the stars themselves, to explain the systematic differences among associations and clusters discussed in the previous section.  However, this behavior is quite different from that of 
ON vs.\ OC spectra (Walborn 1976, 2003; Walborn \& Howarth 2000), which
show opposite anomalies in N vs.\ C, O that correlate with the He abundance 
and certainly result from differing degrees of mixing or binary transfer of 
CNO-processed material into the atmospheres and winds: absent in the OC, 
present in normal spectra, and extreme in the ON (Smith \& Howarth 1994, 
Maeder \& Conti 1994).  As always, the morphology provides strong hypotheses for subsequent physical investigation.

The Of?p stars are also N-rich (Naz\'e, Walborn, \& Martins 2008).  In these peculiar objects, the phenomenology indicates that the C~III $\lambda\lambda$4647--4650--4652 emission is localized to some particular 
region of the unknown circumstellar structures, which is sharply occulted 
at certain rotational phases.  However, the C~III $\lambda$5696 emission 
remains constant as a function of phase (Walborn et~al.\ 2004b), and thus 
arises in a different location, most likely the stellar atmospheres.  The strong dependence of the $\lambda\lambda$4647--4650--4652 emission on spectral type (and thus, likely on the atmospheric parameters) in the Ofc stars may be related to its extreme spatial localization in the Of?p circumstellar environments.  Thus, a physical explanation of one of these categories may illuminate that of the other.

\acknowledgments
N.R.W.\ thanks the Instituto de Astrof\'{\i}sica de Andaluc\'{\i}a and its 
staff for kind hospitality and subsistence (through Spanish Government grant 
AYA2007-64052) during visits there in 2008 and 2009.  His travel to Spain 
was supported by NASA through grants GO-10898.01 and GO-11212.03 from STScI.  A.S., J.M.A, and E.J.A.\ acknowledge support by the Spanish Government Ministerio de Ciencia e Innovaci\'on grant AYA2007-64052 and by the Junta de Andaluc\'{\i}a grant P08-TIC-4075; J.M.A.\ is also supported by the Ram\'on y Cajal Fellowship program, cofinanced by the European Regional Development Fund (FEDER).  R.H.B.\ acknowledges partial support from Universidad de La Serena Project DIULS CD08102.  J.I.A.\ is supported by European Southern Observatory--Chilean Government Joint Committee funds.  We are grateful for generous allocations of observing time at OSN, CAHA, and LCO, without which a program of this magnitude would not be feasible.  Publication was supported by the STScI Director's Discretionary Research Fund.

\begin{deluxetable}{llllrrcl}
\tabletypesize{\small}
\tablewidth{0pt}
\tablecolumns{8}
\tablecaption{Stellar and Observational Data for Ofc and Of?p Stars}
\tablehead{
\colhead{Name} &\colhead{R.A.(2000.0)} &\colhead{Dec.(2000.0)} 
&\colhead{Sp.Type} &\colhead{$V$} &\colhead{$B-V$} 
&\colhead{Observ.} &\colhead{Date}
}
\startdata
\noalign{\vspace{-2pt}}
\multicolumn{8}{c}{Ofc}\\[3pt]
\tableline
\noalign{\vspace{5pt}}
CPD $-47^{\circ}$ 2963   & 08:57:54.620 & $-$47:44:15.71  & O5 Ifc   &  8.45 
& +0.74  & LCO &2008 May 22\\
Cyg OB2--11           & 20:34:08.514 & +41:36:59.42  & O5.5 Ifc      & 10.03 
& +1.49  & CAHA &2008 Oct 15\\
Cyg OB2--9            & 20:33:10.734 & +41:15:08.25  & O4.5 Ifc      & 10.96 
& +1.81  & CAHA &2008 Oct 14\\
Cyg OB2--8A           & 20:33:15.078 & +41:18:50.51  & O5 III(fc)    &  9.06 
& +1.30  & CAHA &2008 Jun 25\\
Cyg OB2--8C           & 20:33:17.977 & +41:18:31.19  & O4.5 III(fc)  & 10.19 
& +1.24  & CAHA &2008 Jun 25\\
HD 93403             & 10:45:44.122 & $-$59:24:28.15  & O5 III(fc)    &  7.27 
& +0.21  & LCO &2008 May 23\\
HD 93843             & 10:48:37.769 & $-$60:13:25.53  & O5 III(fc)    &  7.31 
& $-$0.03  & LCO &2008 May 23\\
HD 93204             & 10:44:32.336 & $-$59:44:31.00  & O5.5 V((fc))  &  8.44 
& +0.09  & LCO &2008 May 22\\
HDE 303308           & 10:45:05.919 & $-$59:40:05.93  & O4 V((fc))    &  8.16 
& +0.13  & LCO &2008 May 23\\
HDE 319699           & 17:19:30.417 & $-$35:42:36.14  & O5 V((fc))    &  9.62 
& +0.79  & LCO &2008 May 23\\
\cutinhead{Of?p}
NGC 1624--2         & 04:40:37.266 & +50:27:40.96  & O7f?p         & 11.77 
& +0.57  & CAHA &2009 Oct 30\\
HD 148937            & 16:33:52.387 & $-$48:06:40.47  & O6f?p         &  6.72 
& +0.33  & LCO &2008 May 21\\
HD 191612            & 20:09:28.608 & +35:44:01.31  & O6f?p         &  7.77 
& +0.26  & OSN &2009 Jul 1\\
&&&O8fp          &       
&        & OSN &2007 May 28\\
HD 108               & 00:06:03.386 & +63:40:46.75  & O8fp          &  7.38 
& +0.16  & OSN &2009 Sep 4\\
CPD $-28^{\circ}$ 2561   & 07:55:52.854 & $-$28:37:46.78  & O6.5fp  &  9.97 
& +0.18  & LCO &2008 May 23\\
\enddata
\tablecomments{Most recent references to known Ofc binaries: Cyg~OB2--9: van Loo et~al.\ 2008, Naz\'e et~al.\ 2008; Cyg~OB2--8A: De Becker et~al.\ 2004, component types O5.5 and O6; HD~93403: Rauw et~al.\ 2002, component types O5.5 and O7.\\
Ofc association/cluster memberships in addition to Cyg~OB2: Carina Nebula: 
HD~93204, HD~93403, HD~93843, HDE~303308; NGC~6334: HDE 319699.}
\end{deluxetable}

\begin{figure}
\epsscale{0.5}
\plotone{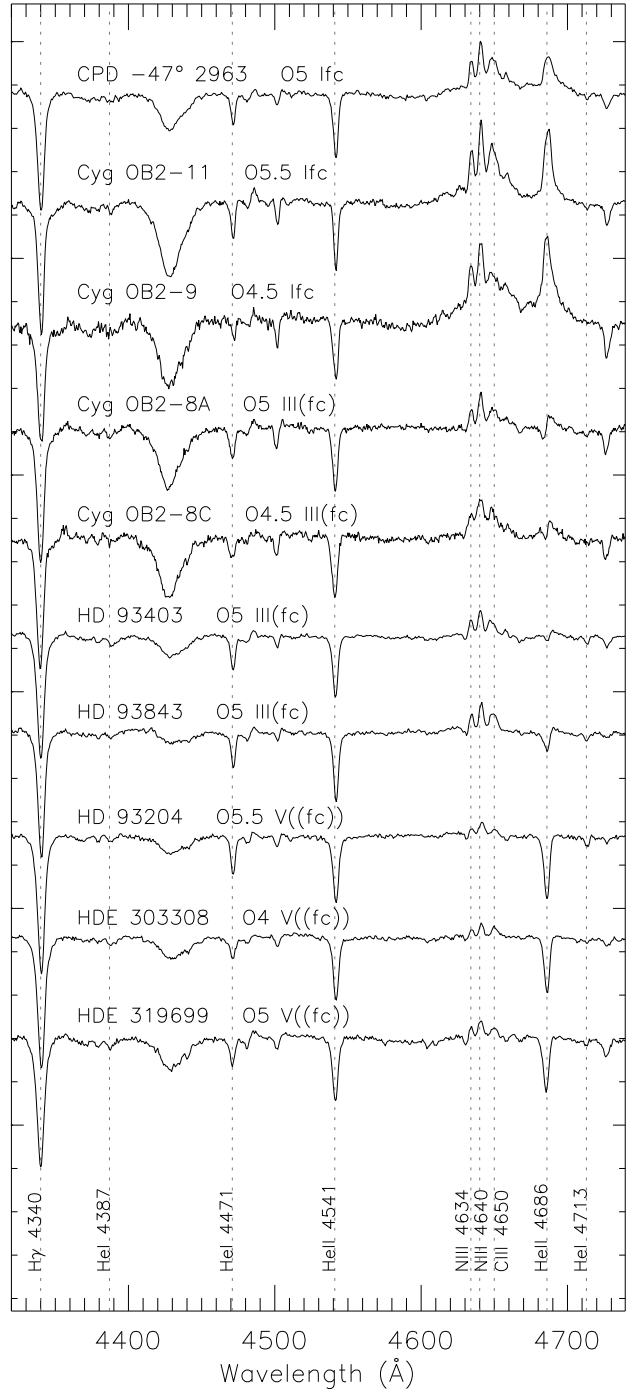}
\caption{\label{fig:fig1}
A luminosity sequence of Ofc spectra in the blue-green region.  The 
ordinate is rectified intensity and the longer tick marks are separated 
by 0.5 continuum units.  In addition to the stellar lines identified, the 
unidentified diffuse interstellar bands at $\lambda\lambda$4430, 4502, 4726 
are prominent.}
\end{figure}

\begin{figure}
\epsscale{0.6}
\plotone{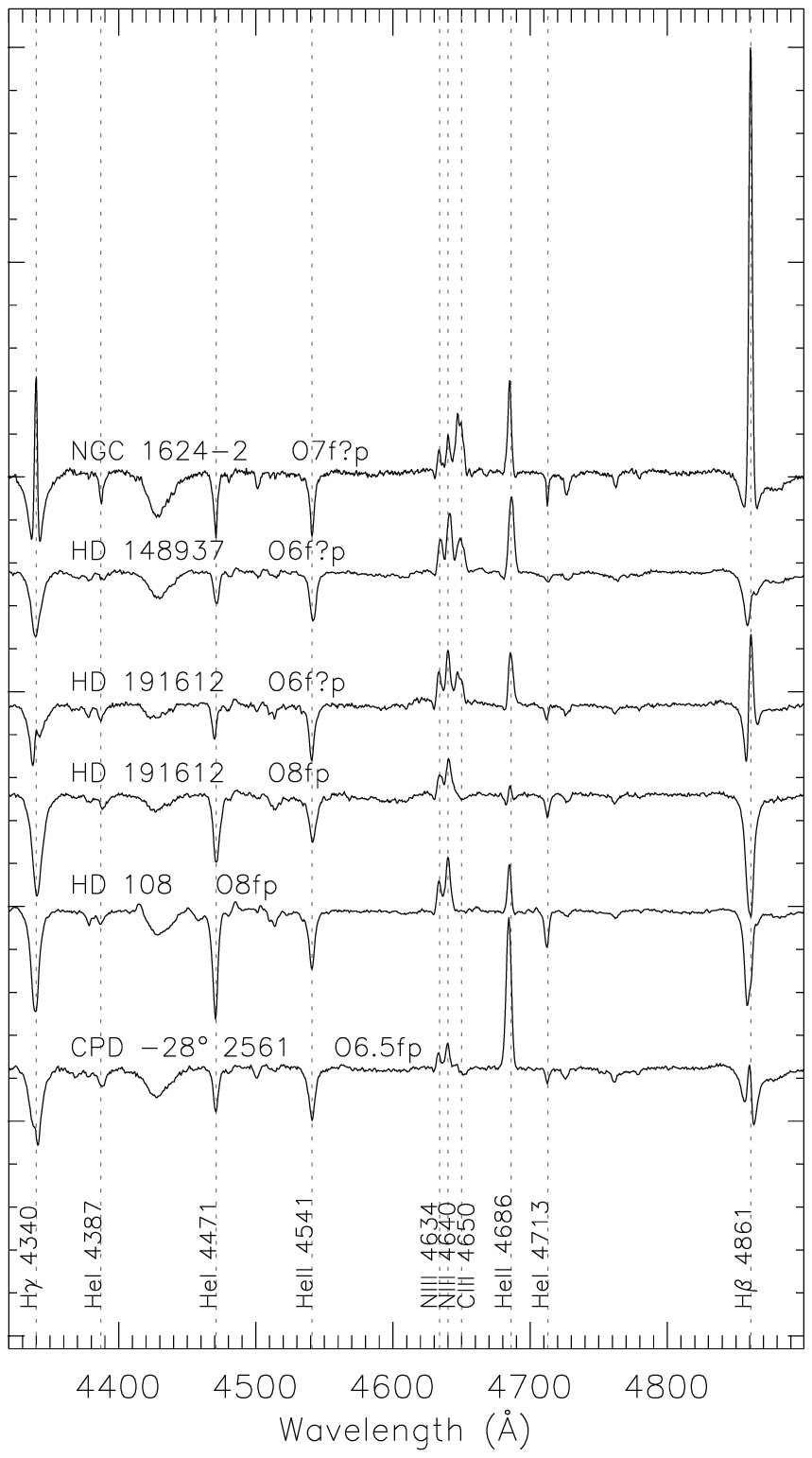}
\caption{\label{fig:fig2}
Spectra of the Galactic Of?p stars in the blue-green.  See the Figure~1 
caption for plot details.}
\end{figure}


\begin{references}

\reference{} Abt, H.A., Meinel, A.B., Morgan, W.W., \& Tapscott, J.W.
1968, An Atlas of Low-Dispersion Grating Stellar Spectra

\reference{} Chini, R., \& Wink, J.E. 1984, \aap, 139, L5

\reference{} Conti, P.S. 1974, \apj, 187, 539

\reference{} Conti, P.S., \& Alschuler, W.R. 1971, \apj, 170, 325

\reference{} Conti, P.S., \& Leep, E.M. 1974, \apj, 193, 113

\reference{} Corti, M.A., Walborn, N.R., \& Evans, C.J. 2009, \pasp, 121, 9 

\reference{} De Becker, M., Rauw, G., \& Manfroid, J. 2004, \aap, 424, L39

\reference{} Deharveng, L., Lefloch, B., Kurtz, S., Nadeau, D.,
Pomar\`es, M., Caplan, J., \& Zavagno, A. 2008, \aap, 482, 585

\reference{} Donati, J.-F., Howarth, I.D., Bouret, J.-C., Petit, P.,
Catala, C., \& Landstreet, J. 2006, \mnras, 365, L6

\reference{} Garrison, R.F., Hiltner, W.A., \& Schild, R.E. 1977, \apjs,
35, 111

\reference{} Heap, S.R. 1977, \apj, 215, 609

\reference{} Herrero, A., Corral, L.J., Villamariz, M.R., \& Mart\'{\i}n, E.L. 
1999, \aap, 348, 542

\reference{} Howarth, I.D., Walborn, N.R., Lennon, D.J., Puls, J.,
Naz\'e, Y. et al. 2007, \mnras, 381, 433

\reference{} Johnson, H.L., \& Morgan, W.W. 1953, \apj, 117, 313

\reference{} Maeder, A., \& Conti, P.S. 1994, \araa, 32, 227

\reference{} Maeder, A., \& Meynet, G. 2000, \araa, 38, 143

\reference{} Ma\'{\i}z Apell\'aniz, J., Walborn, N.R., Galu\'e, H.A., \&
Wei, L.H. 2004, \apjs, 151, 103

\reference{} Martins, F., Bouret, J.-C., Donati, J.-F., \& Marcolino, W.
2010, in preparation

\reference{} Mihalas, D., \& Hummer, D.G. 1973, \apj, 179, 827

\reference{} Mihalas, D., Hummer, D.G., \& Conti, P.S. 1972, \apj, 175, L99

\reference{} Moffat, A.F.J, FitzGerald, M.P., \& Jackson P.D. 1979,
\aaps, 38, 197 

\reference{} Morgan, W.W., \& Keenan P.C. 1973, \araa, 11, 29

\reference{} Naz\'e, Y., De Becker, M., Rauw, G., \& Barbieri, C. 2008, 
\aap, 483, 543

\reference{} Naz\'e, Y., Vreux, J.-M., \& Rauw, G. 2001, \aap, 372, 195

\reference{} Naz\'e, Y., Walborn, N.R., \& Martins, F. 2008, Rev. Mex.
Astron. Astrof., 44, 331

\reference{} Plaskett, J.S., \& Pearce, J.A. 1931, Pub. Dominion Ap. Obs.,
5, 99

\reference{} Rauw, G., Vreux, J.-M., Stevens, I.R., Gosset, E., Sana, H., 
Jamar, C., \& Mason, K.O. 2002, \aap, 388, 552

\reference{} Sharpless, S. 1959, \apjs, 4, 257

\reference{} Smith, K.C., \& Howarth, I.D. 1994, \aap, 290, 868

\reference{} Smith, L.F., \& Aller, L.H. 1969, \apj, 157, 1245

\reference{} Sota, A., Ma\'{\i}z Apell\'aniz, J., Walborn, N.R.,
Alfaro, E.J., Barb\'a, R.H., Morrell, N.I., Gamen, R.C., \& Arias, J.I.  
2010, in preparation

\reference{} Swings, P., \& Struve, O. 1942, \apj, 96, 254

\reference{} van Loo, S., Blomme, R., Dougherty, S.M., \& Runacres, M.C.
2008, \aap, 483, 585

\reference{} Walborn, N.R. 1971a, \apjs, 23, 257

\reference{} ----- 1971b, \apj, 167, L31

\reference{} ----- 1972, \aj, 77, 312

\reference{} ----- 1973a, \aj, 78, 1067

\reference{} ----- 1973b, \apj, 180, L35

\reference{} ----- 1976, \apj, 205, 419

\reference{} ----- 1980, \apjs, 44, 535

\reference{} ----- 2001, in ASP Conf. Ser., 242, Eta Carinae \& Other Mysterious
Stars, ed. T. Gull, S. Johansson, \& K. Davidson (San Francisco: ASP), 217

\reference{} ----- 2003, in ASP Conf. Ser., 304, CNO in the Universe,
ed. C. Charbonnel, D. Schaerer, \& G. Meynet (San Francisco: ASP), 29
   
\reference{} ----- 2009, in Stellar Spectral Classification, ed.
R.O. Gray \& C. Corbally (Princeton University Press), p. 66

\reference{} Walborn, N.R., \& Fitzpatrick, E.L. 1990, \pasp, 102, 379

\reference{} Walborn, N.R., \& Howarth, I.D. 2000, \pasp, 112, 1446

\reference{} Walborn, N.R., Howarth, I.D., Herrero, A., \& Lennon, D.J.
2003, \apj, 588, 1025

\reference{} Walborn, N.R., Howarth, I.D., Lennon, D.J., Massey, P.,
Oey, M.S., Moffat, A.F.J., Skalkowski, G., Morrell, N.I., Drissen, L.,
\& Parker, J.Wm. 2002, \aj, 123, 2754

\reference{} Walborn, N.R., Howarth, I.D., Rauw, G., Lennon, D.J.,
Bond, H.E., Negueruela, I., Naz\'e, Y., Corcoran, M.F., Herrero, A., \&
Pellerin, A. 2004b, \apj, 617, L61

\reference{} Walborn, N.R., Morrell, N.I., Howarth, I.D., Crowther, P.A.,
Lennon, D.J., Massey, P., \& Arias, J.I. 2004a, \apj, 608, 1028
   
\reference{} Werner, K., \& Rauch, T. 2001, in ASP Conf. Ser. 242,
Eta Carinae \& Other Mysterious Stars, ed. T. Gull, S. Johansson,
\& K. Davidson (San Francisco: ASP), 229

\end{references}
\end{document}